\documentclass[prl,twocolumn,showpacs,preprintnumbers]{revtex4}
\usepackage{graphicx}
\usepackage{dcolumn}

\begin{document}

\title{One-dimensional resistive states in quasi-two-dimensional superconductors}

\author{M. Bell} \affiliation{Electrical Engineering
Department, University at Buffalo, Buffalo, New York 14260}
\author{A. Sergeev} \affiliation{Electrical Engineering
Department, University at Buffalo, Buffalo, New York 14260}
\author{V. Mitin} \affiliation{Electrical Engineering
Department, University at Buffalo, Buffalo, New York 14260}
\author{J. Bird} \affiliation{Electrical Engineering
Department, University at Buffalo, Buffalo, New York 14260}
\author{A. Verevkin}\email{verevkin@buffalo.edu} \affiliation{Electrical Engineering
Department, University at Buffalo, Buffalo, New York 14260}
\author{G. Gol'tsman} \affiliation{Physics Department, Moscow Pedagogical State University, 119992 Moscow}


\begin{abstract}

We investigate competition between one- and two-dimensional
topological excitations - phase slips and vortices - in formation
of resistive states in quasi-two-dimensional superconductors in a
wide temperature range below the mean-field transition temperature
$T_{C0}$. The widths $w$ = 100 nm of our ultrathin NbN samples is
substantially larger than the Ginzburg-Landau coherence length
$\xi$ = 4nm and the fluctuation resistivity above $T_{C0}$ has a
two-dimensional character. However, our data shows that the
resistivity below $T_{C0}$ is produced by one-dimensional
excitations, - thermally activated phase slip strips (PSSs)
overlapping the sample cross-section. We also determine the
scaling phase diagram, which shows that even in wider samples the
PSS contribution dominates over vortices in a substantial region
of current/temperature variations. Measuring the resistivity
within seven orders of magnitude, we find that the quantum phase
slips can  only be essential below this level.
\end{abstract}

\maketitle The nature of the resistive state in superconductors
attracts much attention from the physics community, since it
involves fundamental phenomena and advanced concepts \cite{1,2}
such as the mechanisms of high-$T_c$ superconductivity \cite{3},
thermal fluctuations \cite{4,5}, macroscopic quantum tunneling
\cite{6,7,8}, coherence \cite{9}, topological excitations
\cite{10}, and phase disordering \cite{11}. Resistive states are
used in a number of quantum nanodevices, such as logic elements
\cite{12}, ultra-sensitive detectors of radiation, single-photon
counters, and  nanocalorimeters \cite{13,14}. Understanding
resistive states in nanoscale superconductors is critical for the
advancement of fundamental science and the development of novel
applications.

Phase slips and vortices are elementary topological excitations
which create resistive states \cite{1,2,10}. Wires with radius
less than the coherence length $\xi$, or stripes with $w < \xi$
are one-dimensional (1D) superconductors. In 1D-structures, the
resistive state is produced by phase slips and is well described
by the Langer, Ambegaokar, McCumber, and Halperin (LAMH) theory of
thermally activated phase slips (TAPSs) \cite{1,2,4,5}. At low
enough temperatures, the quantum phase slips (QPSs) should be
important \cite{6}, but the magnitude of this effect and the
characteristic resistance at the transition from TAPSs to QPSs are
still under debate \cite{7,8}.

In 2D-superconductors the resistive state is formed by moving
vortices. Above the Berezinskii-Kosterlitz-Thouless (BKT)
transition temperature, $T_C$, there is a nonzero concentration of
free vortices due to thermal unbinding of vortex-antivortex pairs
(VAP) \cite{15,16}. Below $T_C$ in an infinite 2D-superconductor,
VAPs are tightly bound and only a significant bias current can
unbind the pairs, resulting in a nonlinear flux flow resistance.
In finite size samples, free vortices can exist below $T_C$ and
produce a linear resistance at low bias currents \cite{15}. It is
commonly believed that the transition from 1D phase slip
excitations to 2D vortex physics takes place at $w/\xi \sim 1$
\cite{1,2,7,12,17}. However, despite thorough studies of phase
slip and vortex mechanisms, an investigation of their competition
in quasi-two-dimensional superconductors is long overdue.

In this paper we study the resistive transition in NbN
superconducting samples of thickness $d$ = 4nm, width $w$ = 100
nm, and a relatively short coherence length, $\xi \simeq $ 4 nm,
primarily due to the short electron mean free path. Thus, our
samples are quasi-2D superconductors with a ratio $w/\xi$ = 25. We
show that in these samples, 1D phase slip strips (PSSs) are
responsible for resistive state formation in a wide temperature
range, extending far below the BKT transition. We also determine
the phase diagram and show its scaling character.

Our structures were fabricated from NbN superconducting films,
which were deposited on R-plane sapphire substrates by DC reactive
magnetron sputtering. Details of the film deposition process are
described in \cite{18}. The samples were then patterned using
direct electron beam lithography and reactive ion etching.
Parameters for each of the samples are shown in Table I.

Transport measurements were performed in a vacuum. 4-point
resistance measurements with RF-filtered leads were made with a
lock-in amplifier for bias currents ranging from 10 to 300 nA. A
4-point DC measurement setup with RF-filtered leads was used for
higher currents ranging from 500 nA to 1 $\mu$A. In Fig.1 we
present the resistance vs. temperature obtained for both samples
S1 (Fig. 1.a) and S2 (Fig. 1.b) at a bias current of 10 nA.

We begin by investigating the resistivity fluctuation region right
above the mean field superconducting transition $T_{C0}$. The
insets in Figs. 1.a and 1.b show the resistivity on a linear
scale, just above the superconducting transition for both samples.
Our analysis shows that the data obtained is well described by the
Aslamazov-Larkin (AL) fluctuation conductivity \cite{19},
\begin{eqnarray}
\sigma_{AL}= {e^2\over 16 \hbar d}
 {T_{C0}\over T-T_{C0}}.
\end{eqnarray}
Fitting the change in resistivity by $\delta \rho(T) =-
\sigma_{AL} \rho_N^2$, we aim to determine $T_{C0}$ and $\rho_N$.
The obtained parameters are presented in Table I. Note that we do
not study the corrections very close to $T_{C0}$ and limit the
range of data fitted to relatively small corrections, $\delta \rho
/\rho_N \leq 0.1$. Our fits in the insets of Fig. 1 show that in
this temperature range the AL term in the form of Eq. 1 dominates
over other possible contributions.

Next, we investigate the resistivity below $T_{C0}$ in the
framework of the LAMH theory for TAPSs. The LAMH expression for
resistance at low bias currents is \cite{1,4,5}
\begin{eqnarray}\label{1,2}
R_{LAMH}(T) &=& {4 R_0 L\over \pi \xi(T)} {T_{C0}-T\over T_{C0}}
\sqrt{\Delta F \over k_B T} \exp{\biggl(-{\Delta F \over k_B
T}\biggr)}, \ \ \\ \Delta F(T) &\approx& 0.4 k_B(T_{C0}-T) {w\over
\xi(T)} {R_0 \over R_{sq}},
\end{eqnarray}
where $R_0= h/2e^2 \simeq 13$ k$\Omega$  is the resistance
quantum, $\xi(T)=\xi(1-T/T_{C)})^{-1/2}$ is the temperature
dependent coherence length, and $R_{sq}= \rho_N/d$ is the normal
state resistance per square.

In Fig. 1 we show fits of our $R(T)$ data to LAMH formulas (Eqs. 2
and 3), which are in excellent agreement for both samples. In this
fitting procedure we used only one fitting parameter $\xi$, which
was found to be 4.0 nm for both samples. This value is in a good
agreement with $\xi \approx 0.6 \sqrt{\hbar D/k_BT_{C0}}$ = 3.4
nm, determined with the diffusion coefficient $D \simeq $0.5
cm$^2$/s, which can be obtained from the resistivity of our films
\cite{20}.

Measuring the linear resistance as a function of temperature in a
range of seven orders of magnitude, we do not observe any
deviation from LAMH theory, which is usually considered as a
manifestation of QPSs. Although several groups reported
observations of QPSs \cite{6,7,8}, the data and its interpretation
remain controversial.  While various phenomenological and
microscopic models of QPS formation have been proposed,
experimentalists appeal to the phenomenological theory suggested
by Giordano \cite{6}. In this approach, the QPS contribution to
the resistivity can be described by Eq. 2, where $k_B T$ is
replaced by $(8a k_B/\pi) (T_{C0}-T)$, with $a$ a constant of the
order of unity. Also, in this phenomenology an additional fitting
parameter $b$ was added into $\Delta F (T)$. According to
Giordano, $a$ is a universal constant, so the crossover from TAPS
to QPSs should occur at a characteristic value of
$t_{cr}=T/T_{C0}$ independent on the sample cross-section and
material parameters. The factor $b$ is weakly dependent on
material properties and, therefore, the crossover should be
observed at approximately the same value of
$r_{cr}=R(t_{cr})/R_N$. In fact, for the first time the crossover
was observed in In wires at $ r_{cr} \simeq 10^{-2} - 10^{-3}$
\cite{6}; then in Sn wires at $r_{cr} \simeq 10^{-3}-10^{-4}$
\cite{7}, and recently in Al wires at $r_{cr} \simeq
10^{-4}-10^{-5}$ \cite{8}. All data shows a strong dependence of
$r_{cr}$ on the wire diameter. Other recent measurements in MoGe
wires do not show any trace of QPS for $R/R_N$ spanning 11 orders
of magnitude \cite{20}. Our data supports the results of Ref.
\cite{20} and shows that QPSs could be important at lower
temperatures and $ r_{cr}$ values.

We have also investigated nonlinear effects at bias currents
ranging from 10 nA to 1 $\mu$A. The current-voltage
characteristics obtained at various temperatures are shown in Fig.
2. The solid lines represent the LAMH expression \cite{1},
\begin{eqnarray}\label{4}
V(I,T) = 0.5 \ I_T R_{LAMH}(T)  \exp{(I/I_T)},
\end{eqnarray}
where $I_T = 4ek_B T/h \simeq 0.013 \mu {\rm A}/{\rm K} T$. To
plot $V(I,T)$ we used the resistance $R_{LAMH}$ (Eq. 2)
investigated previously at low currents. The dashed line in Fig. 2
represents the characteristic current $I_T$. As shown in Fig. 2,
the results of our measurements agree with the LAMH theory for
currents up to $\sim$ 1 $\mu$A. Deviations at higher currents are
most likely due to electron heating, which will be evaluated
later. Note that, in thin-film quasi-2D superconductors, the
heating can be localized into resistive domains, which are formed
by the dissipative motion of vortices \cite{9}; however, our data
does not show such effects of local heating.

We have shown above that our data on resistivity measured in ohmic
and non-ohmic regimes are in excellent agreement with the LAMH
theory, which has been developed for quasi-1D superconductors. Why
the LAMH theory turns out to be applicable to quasi-2D samples?
Generally speaking, in superconductors with width wider than
$\xi$, the order parameter can change along the width and,
therefore, resistive domains of various geometries are possible.
Eqs. 2 and 3 take into account only simple domains in the form of
PSSs, i.e. strips with width equal to $\xi $ and a length equal to
$w$ (Fig. 3.a). Obviously, among other possible forms, the PSS has
a minimal volume (see Fig. 3) and its generation requires a
minimal energy, $\Delta F \simeq (H_c^2/8\pi) \xi w d $, where
$H_c$ is the critical magnetic field. Because of the exponential
dependence of $R_{LAMH}$ on $\Delta F$, one can neglect domains
other than PSSs.

Overall, analysis of our data in the framework of the LAMH theory
strongly suggests that the resistive state in our quasi-2D NbN
samples is a result of 1D PSS excitations. This observation is
very interesting, because in such samples vortices are expected to
dominate the resistive state. In what follows we support the
statement above by modeling the competition between 1D PSS
excitations and vortices in quasi-2D superconductors.

The vortex state critically depends on temperature with respect to
the BKT transition temperature $T_C$, introduced for an infinitely
wide film. Above $T_C$, there is a finite concentration of free
vortices due to thermal fluctuations, while below $T_C$ all
vortices are tightly bound into pairs. For disordered
superconducting films, $T_C$ is given by ${T_C /T_{C0}} =
(1+0.54{R_{sq}/R_0})^{-1}$ [21]. We estimate $T_C$ = 11.3 K for
sample S1 and $T_C$ = 11.6 K for sample S2. Since the obtained
values of $T_C$ are very close to $T_{C0}$, and the temperature
interval $T_{C0} - T_{C}$ is quite narrow compared to the entire
resistive transition broadening in our samples, we limit our
calculations to temperatures below $T_C$. In this temperature
range, current-induced unbinding of VAPs can occur when the
Lorentz force proportional to the bias current density exceeds the
force of mutual attraction between the vortex and antivortex. The
threshold current which unbinds the vortex and antivortex
independent of film width is given by \cite{23}
\begin{eqnarray}\label{5}
I_{th} \simeq 2 \kappa(T)I_T, \ \ \kappa(T)={T_{C}-T \over
T_{C0}-T_C}.
\end{eqnarray}

For $I \geq I_{th}$, the nonlinear resistance due to
current-induced unbinding of vortex-antivortex pairs can be
approximated by
\begin{eqnarray}\label{6}
R_{VAP}^{nl} \simeq 4 R_N [\kappa(T)-1] (I/I_0)^{2\kappa(T)},
\end{eqnarray} where $I_0=[w/\xi(T)] I_T$   is a
characteristic current proportional to the structure width.

For $I \leq I_{th}$ , the bias current is not sufficient to unbind
VAPs, but a nonzero concentration of free vortices can still exist
due to the finite size of the film \cite{16}. These free vortices
produce a linear resistance,
\begin{eqnarray}\label{7}
R_{VAP}^{l} \simeq 4 R_N [\kappa(T)-1] \biggl({\xi(T)\over
w}\biggr)^{2\kappa(T)}.
\end{eqnarray}

To investigate PSS and vortex contributions to the resistance, we
use Eqs. 2, 4, 6, and 7. Taking into account that the normal
resistance may be represented as $R_N=R_{sq}(L/\xi)(\xi/w)$, we
see that the relative contributions of PSS and vortices depend on
four dimensionless parameters: $I/I_T$, $(T_C -T)/(T_{C0} -T_C )$,
$w/\xi$, and $R_{sq}/R_0$. In Fig. 4 we present a phase diagram of
the resistive states in $I/I_T$ and $(T_C -T)/(T_{C0} -T_C )$
coordinates for samples with $w/\xi$ = 25 (Fig. 4.a) and $w/\xi$ =
75 (Fig. 4.b). The value of $R_{sq}/R_0$ was taken to be 0.05,
which corresponds to our NbN samples. The solid lines in Fig. 4
represent the boundary between PSS and vortex mechanisms,
described by Eqs. 2 and 4 and Eqs. 6 and 7 correspondingly. The
solid gray lines represent $I_T$ which separate the linear and
nonlinear mechanisms related to PSSs. The nonlinear vortex
mechanism dominates over others at currents, which exceed the
threshold current given by Eq. 5. The hashed area represents a
region of strong electron heating, where the change in electron
temperature $\delta \theta \simeq j^2 \rho \tau_{e-ph}/C_e$ rises
to ~$(T_C -T)$. In these evaluations we have used values for the
electron heat capacity, $C_e$, and the electron-phonon relaxation
time $ \tau_{e-ph}$ from Ref. \cite{23}. Finally, the dashed
segments of the boundaries in both phase diagrams represent a
narrow region very close enough to $T_{C0}$, where the interaction
between phase slips and nonequilibrium phenomena become important
and, therefore, the LAMH theory is not applicable \cite{1}.

As the sample width increases, the relative contribution of PSSs
to the resistivity decreases. The phase diagram for samples with
$w/\xi$ = 75 in Fig. 4.b shows that, at currents below $I_{th}$,
the PSS mechanism competes with free vortices generated below
$T_C$ due to finite size effects. While the contributions from
both mechanisms decrease with increasing sample width, the PSS
formation is more sensitive to the parameter $w/\xi$ because of an
exponential dependence of the PSS resistance (Eq. 2) on $\Delta
F(T)$, which, in turn, is proportional to $w/\xi$. Fig. 5 shows
that the characteristic crossover temperature $T^*$ separating the
linear PSS and vortex mechanisms strongly depends on the sample
width and changes very weakly with $R_{sq}$. Therefore, the
obtained diagram has a scaling character in the dimensionless
coordinates used.

Reviewing other experiments with NbN, let us note that the complex
investigations of resistivity in 1 $\mu$m films \cite{24},
demonstrate only effects of vortices. This result is in agreement
with our modeling, which also does not show any measurable
contribution of PSSs in the samples with $w/\xi = 250$. In
\cite{25} samples with geometry similar to our device S2 and
$T_{C0}$=4.8 K have been investigated at rather high currents,
$\sim$ 1 $\mu$A. The data of Ref. 25 have been found to be well
described by the vortex-induced resistivity. This observation is
also in qualitative agreement with our phase diagram, while
essential effects of electron heating may also be expected.
According to our modeling, samples with $w/\xi$ in the range of 50
- 100 (0.2 -0.4 $\mu$m for NbN) would be preferable for further
experimental investigations of the boundaries between 1D and 2D
excitations.

In conclusion, we have investigated the competition between the
PSS and vortex excitations in the formation of resistive states in
quasi-2D superconductors.  Our data is in excellent agreement with
the LAMH theory of TAPSs and does not show any evidence of QPS
contribution. The data and modeling show that the 1D PSS mechanism
dominates over the substantial range of parameters (see Fig. 4).
Essentially, the creation of a PSS overlapping the sample
cross-section requires a significantly higher energy than the
creation of a vortex or VAP. However, the PSS mechanism turns out
to be more effective in producing resistance. In clean
superconductors with large values of $\xi$, the PSS excitations
will prevail over vortices in wide samples, because the phase
diagram depends only on the parameter $w/\xi$.

The work was supported by NYSTAR. MB also acknowledges support
from the NSF IGERT program.

\begin{table*}
\caption{Sample parameters: the sample thickness ($d$), length
($L$), and width ($w$); the mean field transition temperature
($T_{C0}$), the normal resistivity ($\rho_N$), and the
Ginzburg-Landau coherence length ($\xi$). $T_{C0}$ and $\rho_N$
were determined from the AL conductivity above $T_{C0}$ , $\xi $
is found from the LAMH fit below $T_{C0}$.}
\begin{ruledtabular}
\begin{tabular}{ccccccc}
 Sample & $ d$ & $L$ & $ w $ &$T_{C0}$ & $\rho_N$ & $\xi$ \\
 & nm & $\mu$m &  nm  & K & $\mu \Omega \cdot $cm &  nm \\[3pt] S1
& $ 4 \pm 0.2$ & $500$ & $ 100 \pm 5 $ & $11.6$ & $264$ & $4.0$
\\ S2 & $4 \pm 0.2 $ & $30$ & $ 100 \pm 5 $ &$ 11.9$ &
$252$ & $4.0$ \\
\end{tabular}
\end{ruledtabular}
\end{table*}

\clearpage

\begin{figure}
\caption{Normalized resistance $R/R_N$ vs. $T$ for the samples S1
(a) and S2 (b), the solid lines represent fits by the LAMH theory.
Insets show the resistance above $T_{C0}$, and solid lines are the
AL fits.}
\end{figure}

\begin{figure}
\caption{Current-voltage characteristics of the sample S1 at $T$ =
9.7 K - 10.6 K. The solid lines are fits by the LAMH theory (Eq.
4). The dashed line shows the characteristic current $I_T$.}
\end{figure}

\begin{figure}
\caption{Among other possible forms of resistive domains in
quasi-2D superconductors, the PSS stripe (a) with width equal to
$\xi $ and a length equal to $w$ has a minimal volume.}
\end{figure}

\begin{figure}
\caption{The phase diagrams for samples with $w/\xi$ of 25 and 75.
Hashed pattern represents the area of significant electron
heating.}
\end{figure}

\begin{figure}
\caption{The crossover temperature $T^*$ as a function of sample
width for samples with different sheet resistances.}
\end{figure}

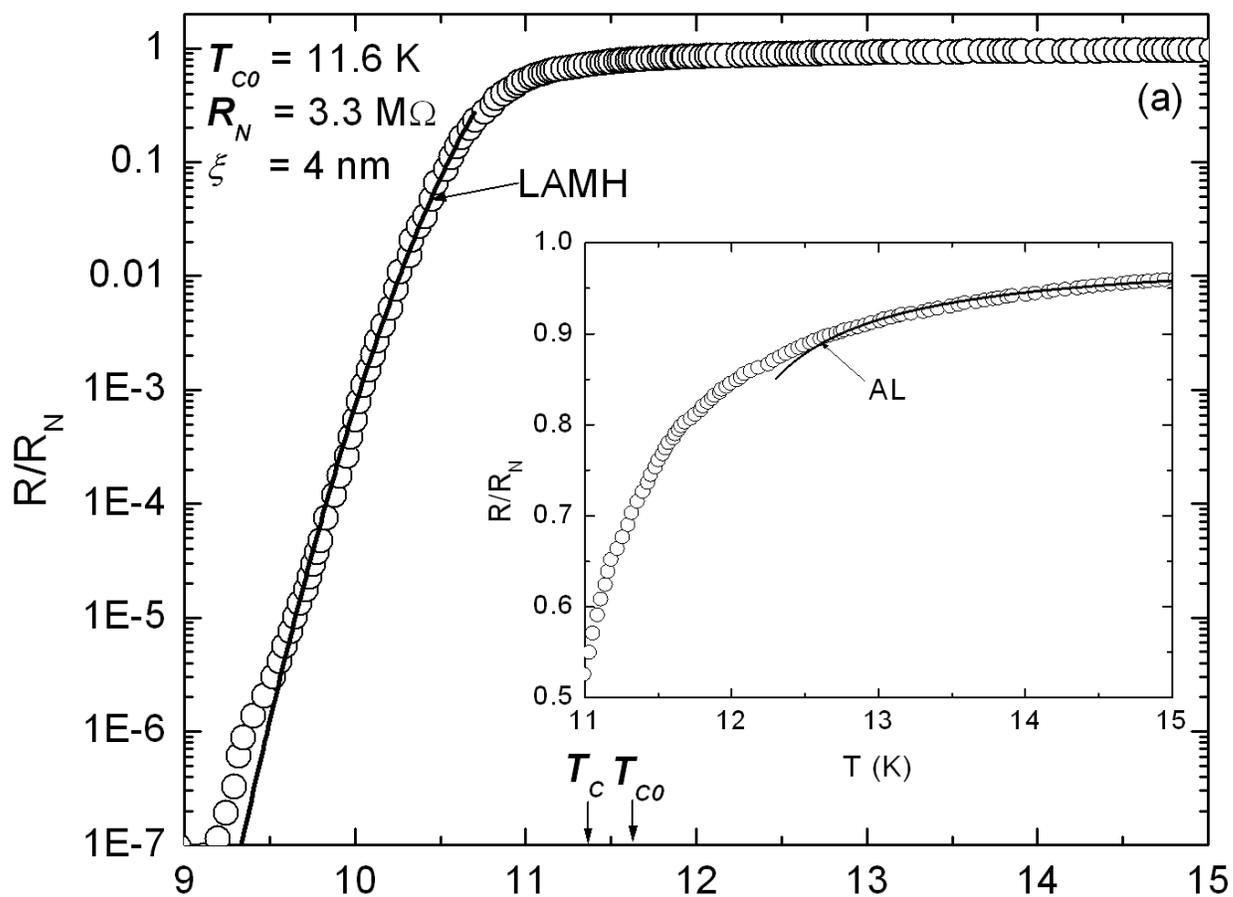
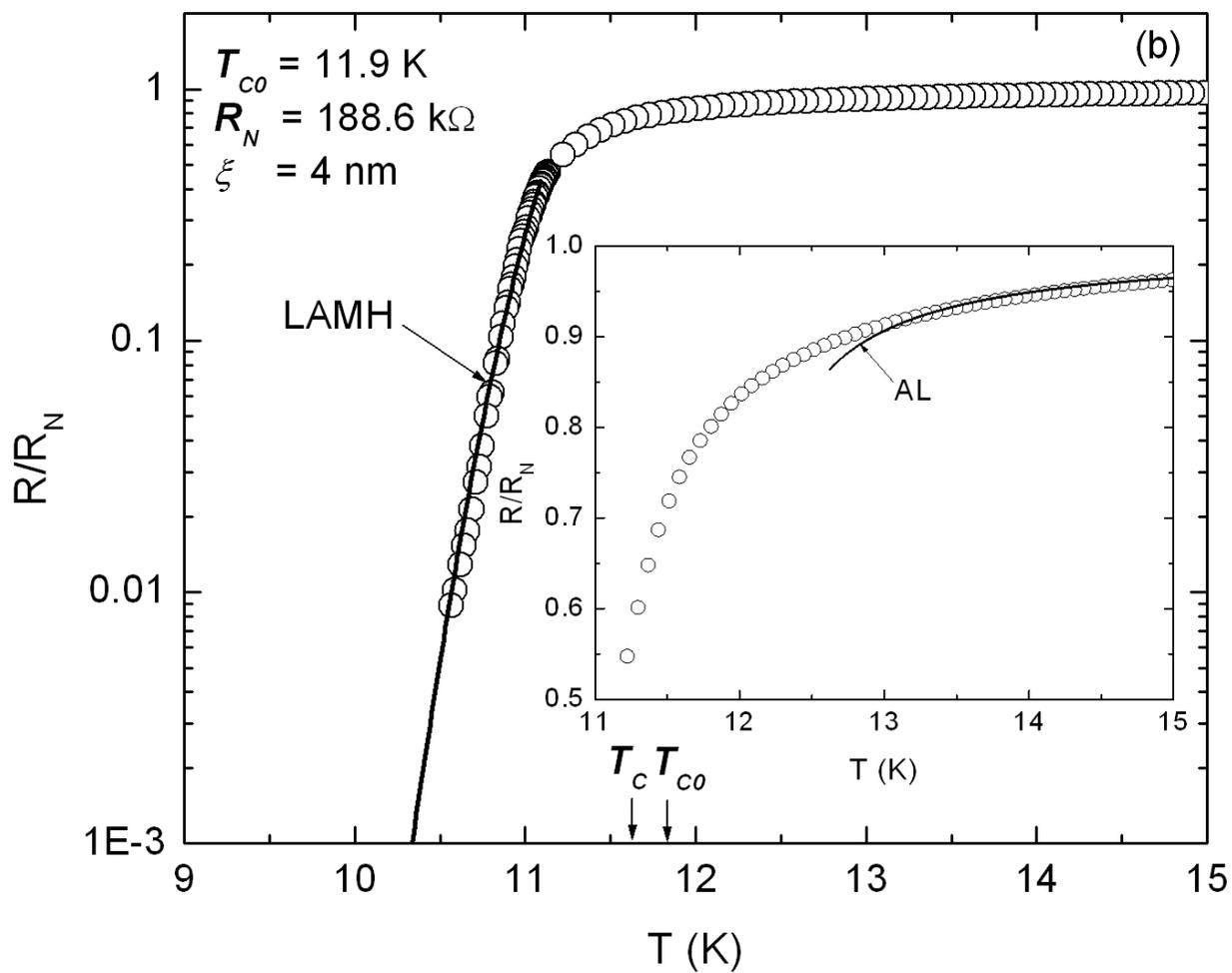
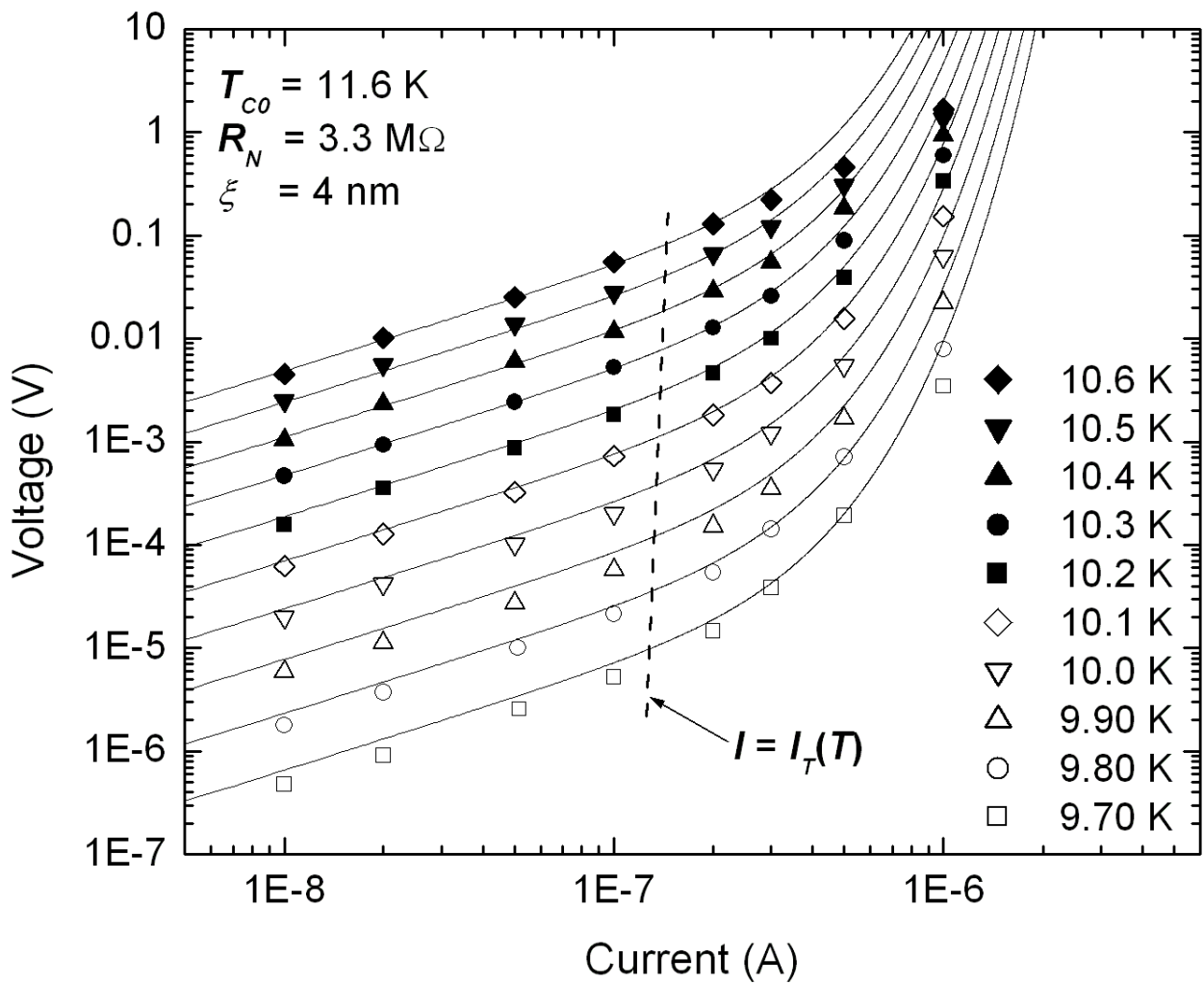
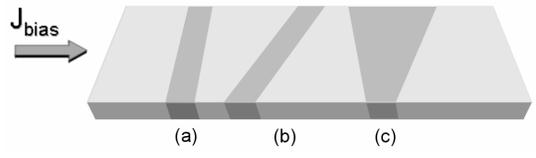
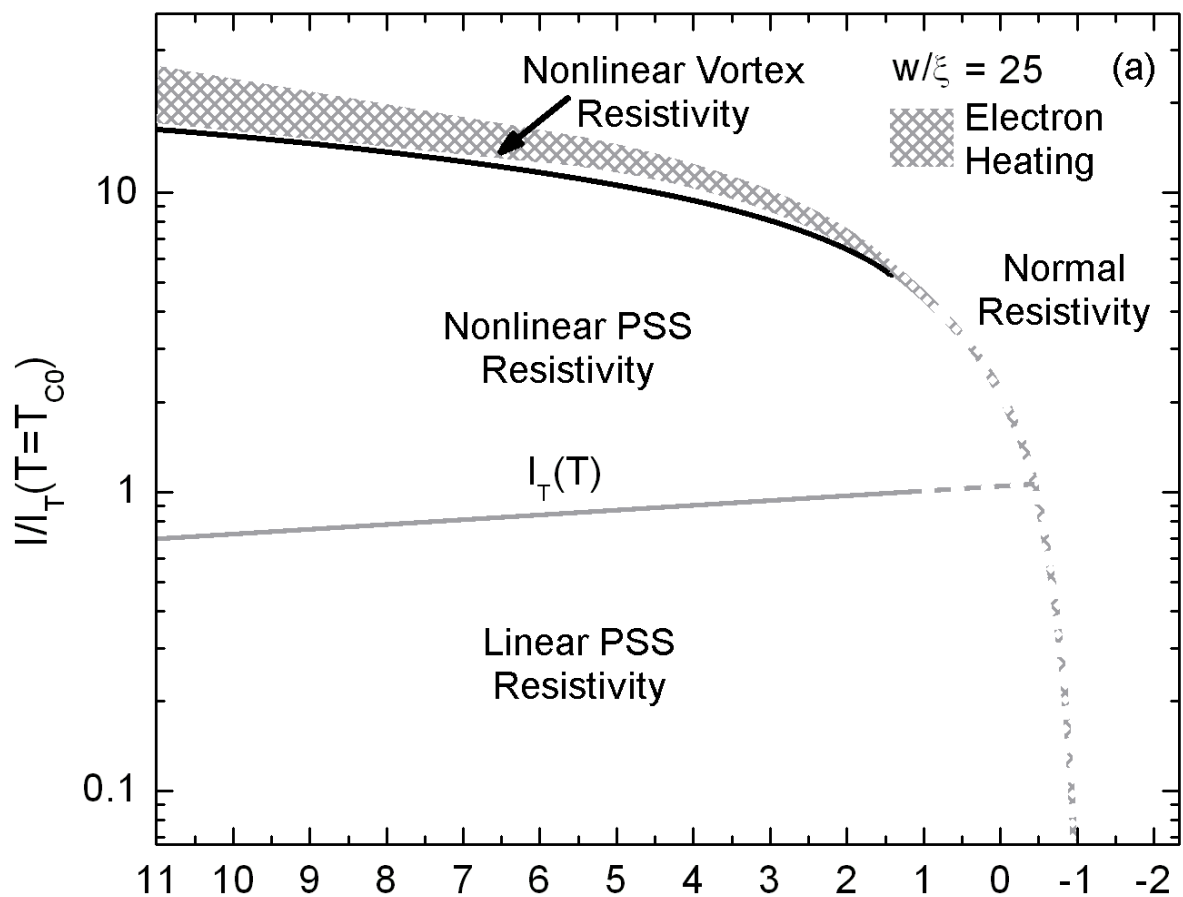
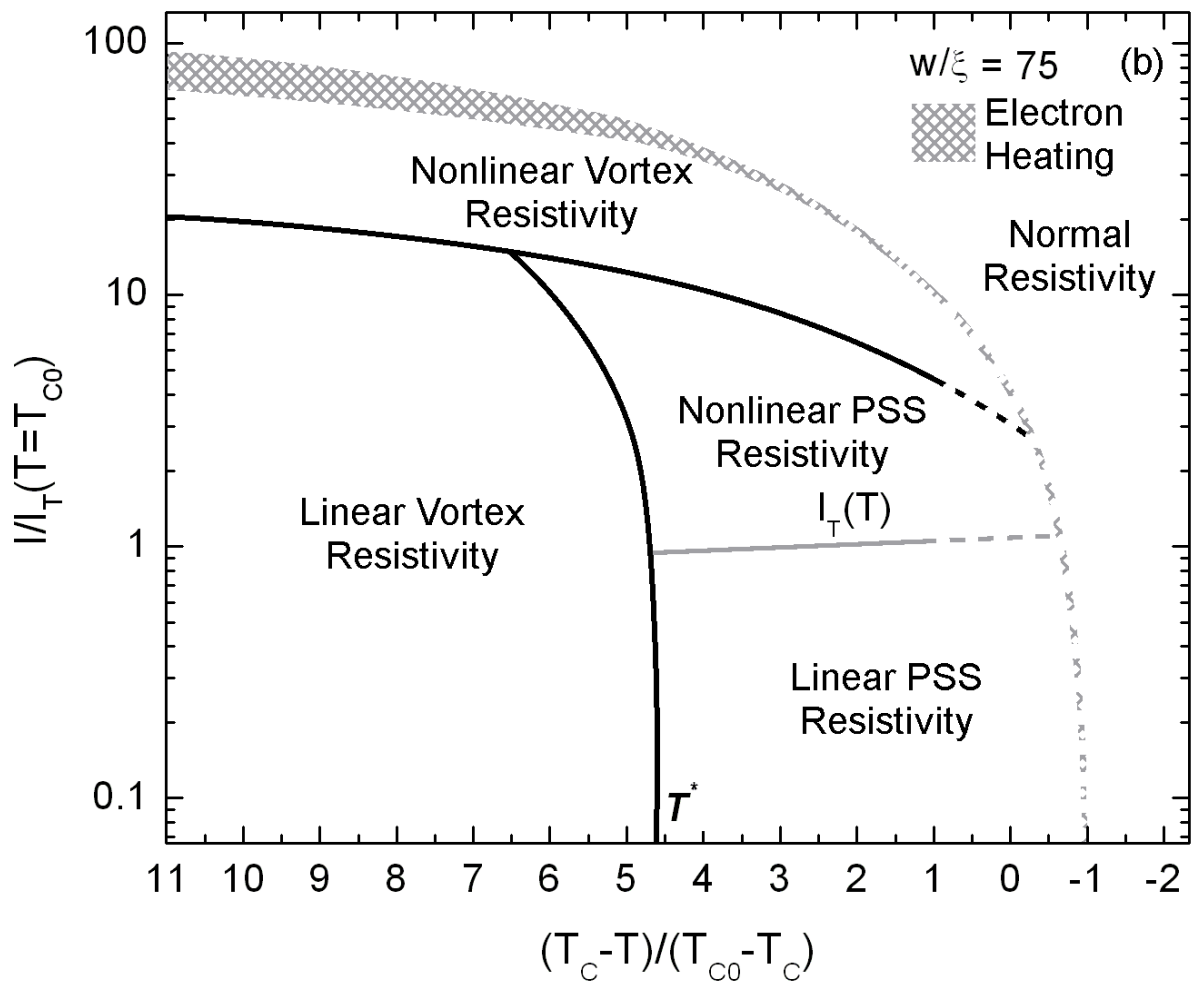
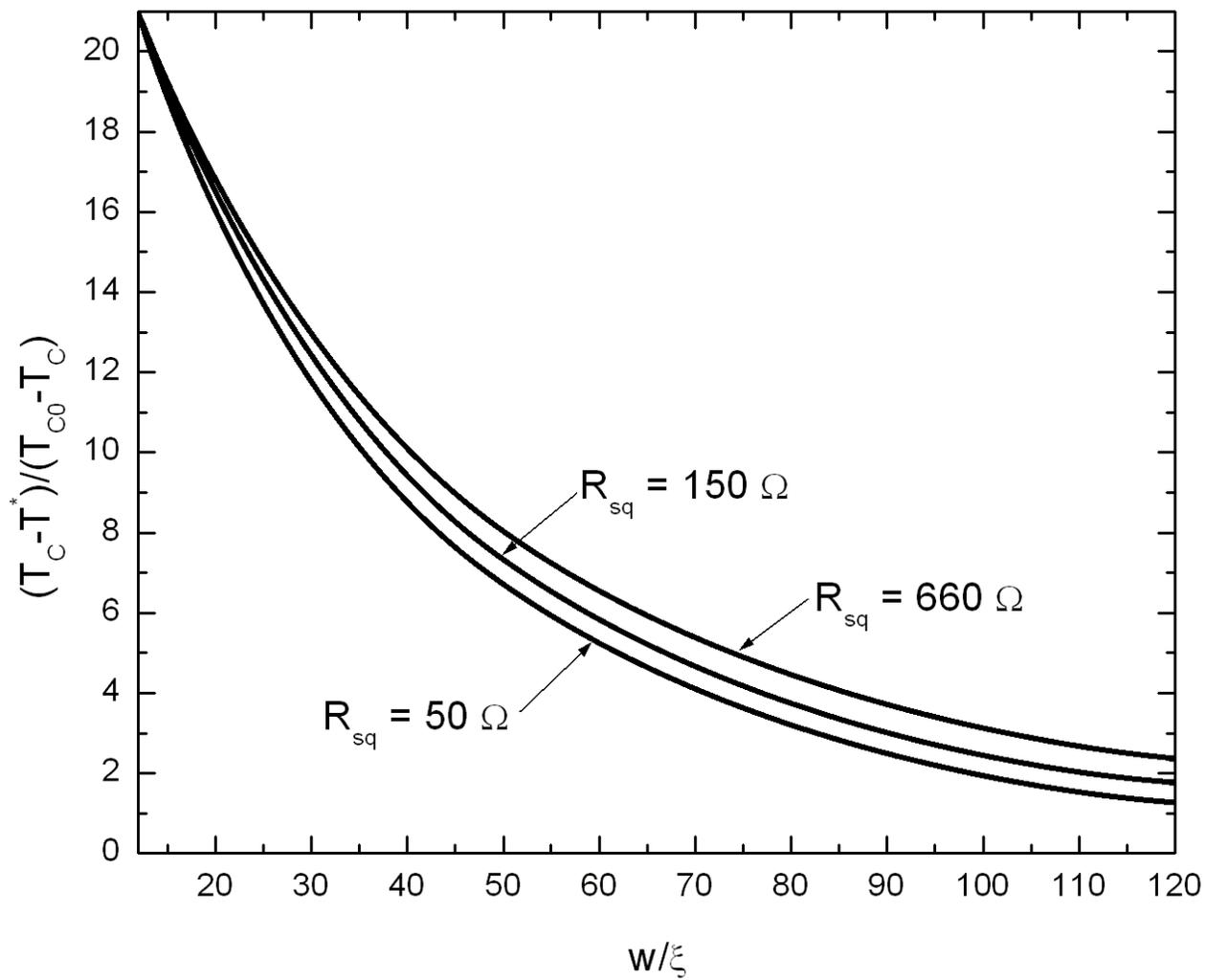
\end{document}